\documentclass[a4paper,12pt]{article}
\usepackage{graphicx}
\usepackage{epstopdf}
\newcommand{\resection}[1]{\setcounter{equation}{0}\section{#1}}
\newcommand{\appsectio}{\setcounter{section}{0}
         \addtocounter{section}{1} \setcounter{equation}{0}
                         \section*{Appendix \Alph{section}}}
\newcommand{\appsection}{\addtocounter{section}{1} \setcounter{equation}{0}
                         \section*{Appendix \Alph{section}}}
\renewcommand{\theequation}{\thesection.\arabic{equation}}
\begin{document}
%%%%%%%%%%%%
\begin{figure}[t]
\begin{center}
\includegraphics[width=7.5cm,angle=0]{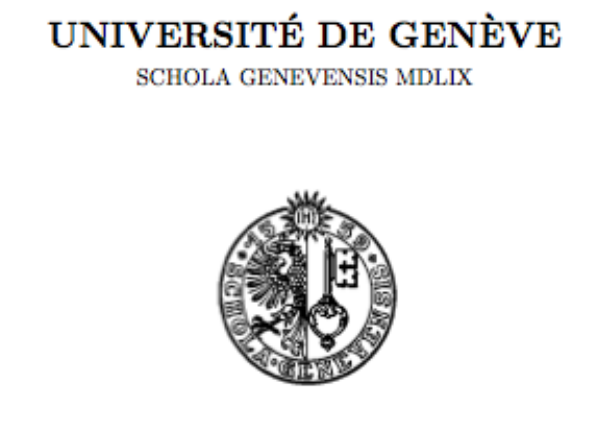}
\end{center}
\end{figure}
%%%%%%%%%%%%

\begin{center}
  \begin{Large}
  \begin{bf}
THE EXTENDED BESS MODEL: BOUNDS FROM PRECISION ELECTROWEAK MEASUREMENTS$^*$\\
  \end{bf}
  \end{Large}
  \vspace{5mm}
  \begin{large}
R. Casalbuoni and S. De Curtis\\
  \end{large}
Dipartimento di Fisica, Univ. di Firenze\\
I.N.F.N., Sezione di Firenze\\
  \vspace{5mm}
  \begin{large}
A. Deandrea, N. Di Bartolomeo and R. Gatto\\
  \end{large}
D\'epartement de Physique Th\'eorique, Univ. de Gen\`eve\\
  \vspace{5mm}
  \begin{large}
D. Dominici\\
  \end{large}
Dipartimento di Matematica e Fisica, Univ. di Camerino, I.N.F.N., Sezione di Firenze\\
  \vspace{5mm}
  \begin{large}
F. Feruglio\\
  \end{large}
Dipartimento di Fisica, Univ.
di Padova, I.N.F.N., Sezione di Padova\\
  \vspace{5mm}
\end{center}
  \vspace{2cm}
\begin{center}
UGVA-DPT 1992/07-778\\\
July 1992
\end{center}
\noindent
$^*$ Partially supported by the Swiss National Foundation
\newpage
\thispagestyle{empty}
\begin{quotation}
\vspace*{5cm}
\begin{center}
  \begin{bf}
  ABSTRACT
  \end{bf}
\end{center}
  \vspace{5mm}
\noindent
We present an effective Lagrangian parameterization describing scalar, vector,
and axial-vector bound states, originating from a strong breaking of the
electroweak symmetry, based on the global symmetry $SU(N)_L\otimes SU(N)_R$.
In this approach vector and axial-vector bound states are gauge bosons
associated to a hidden $SU(N)_L\otimes SU(N)_R$ symmetry. After the
gauging of the electroweak symmetry, the corrections to the self-energies of
the standard model  gauge bosons are calculated and bounds on the parameter
space of the model arising from precision measurements are studied. The
self-energy corrections arise from spin 1 mixings, pseudogoldstones loops,
pseudogoldstone-spin 1 loops, and tadpole terms. The one-loop terms tend to
decrease both isospin conserving and isospin violating corrections. Careful
calculation for standard $SU(8)$ QCD-scaled technicolor shows that strictly
this model (which has however serious theoretical difficulties on his own) is
still marginally allowed at present experimental precision.
\end{quotation}
\newpage
\setcounter{page}{1}
\def\lq{\left [}
\def\rq{\right ]}
\def\qq{Q^2}
\def\dmu{\partial_{\mu}}
\def\dmus{\partial^{\mu}}
\def\AA{{\cal A}}
\def\BB{{\cal B}}
\def\Tr{{\rm Tr}}
\def\gp{g'}
\def\gs{g''}
\def\ggs{\frac{g}{\gs}}
\def\mpp{m_{P^+}}
\def\mpm{m_{P^-}}
\def\mpt{m_{P^3}}
\def\mpz{m_{P^0}}
\def\eps{{\epsilon}}
\newcommand{\be}{\begin{equation}}
\newcommand{\ee}{\end{equation}}
\newcommand{\bea}{\begin{eqnarray}}
\newcommand{\eea}{\end{eqnarray}}
\newcommand{\nn}{\nonumber}
\newcommand{\dd}{\displaystyle}

\resection{Introduction}

The possibility of a new strong interacting sector being at the origin of the
symmetry breaking in the electroweak theory is coming to be quantitatively
testable, after recent precision measurements, particularly at LEP. We shall
deal here with the contributions expected from such strong sector to the vector
boson self-energy corrections.

An expected feature of such strong sector is the occurence of resonances in the
TeV range. The possibility of spin one resonances is particularly
interesting, as they would, already through mixing effects, affect the
self-energies of the standard model gauge bosons. In addition, loop effects,
contributed from both spin-1 and spin-0 particles, whenever they are present,
are also expected to be relatively non negligible and to bear on the
comparison with experiments.

To describe the new strong sector we would like to remain as much as possible
general, without assuming any particular explicit dynamical realization, for
which no definite proposal has been advanced so far. To provide for such
general frame we had developed a model, the BESS model, which was essentially
constructed on the standpoint of custodial symmetry and gauge invariance.

The original BESS was based on the minimal chiral structure $SU(2)_L \times
SU(2)_R$, but it can be easily extended to a larger $SU(N)_L \times
SU(N)_R$ structure. In such a case its most apparent feature is the presence of
spin-0 pseudogoldstones. Extended BESS has been specialized to standard $SU(8)$
technicolor and the conclusion was drawn that the latest experimental data
would exclude conventional QCD-scaled technicolor with $N_{TC}$ technicolors
and $N_{d}$ technidoublets at 90 \% CL for $N_{TC} N_d \geq 12$ \cite{techni}.
It is unnecessary to emphasize that such simple forms of technicolor have
always
had to face theoretical difficulties from their very beginning. This is one
reason why we prefer to go on with the experimental testing of the idea of a
possible strong electroweak sector within a general frame such as BESS, rather
than by adopting any definite dynamical model.

The extended BESS model is based on a chiral global $SU(N)_L \times
SU(N)_R$, and it contains explicit vector and axial-vector resonances (like the
techni-$\rho$ of ordinary technicolor). The phenomenology of ordinary
technicolor would correspond to a specialization of extended BESS.

The simplest construction for extended BESS
uses a local copy of the global chiral symmetry and goes through classification
of the relevant invariants. The same results follow from the hidden gauge
symmetry approach. The standard electroweak $SU(2)\times U(1)$ and $SU(3)$ are
gauged and a definite  mixing scheme emerges for the gauge bosons and the
vector and axial-vector resonances. The physical photon and the physical gluon
remain massless and coupled to their conserved currents.

The quantitative estimates will be restricted to the "historical" case $N=8$,
although a number of results are more general. Through their mixings with the
gauge bosons of $SU(2)_L \times U(1) \times SU(3)_c$, some of the vector and
axial vector resonances acquire a coupling to quarks and leptons, and are thus
expected to be produced at proton-proton and electron-positron colliders of
sufficient high energy. In $SU(8)$ these spin-1 bosons are an $SU(2)$ vector
triplet and axial triplet, an overall singlet, and a vector color octet, the
last one susceptible to be produced through the stronger color interaction.

The effective charged current-current interaction of extended BESS reproduces
the SM interaction, after identification of the relevant scale parameter with
the root of the inverse Fermi coupling. Also, for any chiral $SU(N)_L \otimes
SU(N)_R$, it can be seen that the neutral current-current interaction strength
corresponds to a $\rho$-parameter equal to 1,
because of the diagonal $SU(N)$ which
is supposed to remain unbroken. All these results are of course corrected by
radiative effects.

If one tries to compare $SU(8)$-BESS with the original
$SU(2)$-BESS one sees that one main difference, concerning low energy effective
interaction, lies in the role of the additional singlet vector-resonance,
mentioned above. In addition the extension has new features, notably the
appearance of pseudogoldstones.

In section 2 we present the extended BESS model. In section 3 we evaluate the
tree level corrections arising from BESS to the SM gauge boson self-energies,
whereas the results of the one-loop calculation are given in sections 4 and 5.
Section 6 is devoted to the numerical discussion of our results, which are
further enumerated in section 7. In the appendices are collected some useful
formulas and results.

\resection{The extended BESS model}

To construct the extended BESS model we
proceed analogously to ref. \cite{assiali},
 but starting now from a global symmetry
$SU(N)_L\otimes SU(N)_R$, rather than $SU(2)_L\otimes SU(2)_R$. In order
to do that, one introduces a local copy of the global symmetry,
$[SU(N)_L\otimes SU(N)_R]_{local}$. One  also enforces the idea that
when the new vector and axial-vector particles decouple, one should
obtain the non-linear $\sigma$-model Lagrangian, describing the
Goldstone bosons, transforming as the representation $(N,N)$
of $SU(N)_L\otimes SU(N)_R$, corresponding to the breaking of
$SU(N)_L\otimes SU(N)_R\to SU(N)_{L+R}$
\be
{\cal L} = {v^2\over 2N} Tr \left [(\partial_\mu U)(\partial^\mu U)^\dagger
\right ]
\ee
To introduce both vector and axial-vector particles,
we assume the following factorization of $U$
\be
U=LM^\dagger R^\dagger
\ee
where $L,~M,~R$, transform according to the following representations
of
\be
G=[SU(N)_L\otimes SU(N)_R]_{global}\otimes[SU(N)_L\otimes
SU(N)_R]_{local}\nn
\ee
as
\def\spin{{N}}
\be
L\in (\spin,0,\spin,0)~~~~~
M\in (0,0,\spin,\spin)~~~~~
R\in (0,\spin,0,\spin)
\ee
that is:
\be
L^\prime = g_L L h_L~~~~~
M^\prime = h_R^\dagger M h_L~~~~~
R^\prime = g_R R h_R
\ee
where
\bea
& & g_L\in ({SU(N)_L})_{global}~~~~~
g_R\in ({SU(N)_R})_{global}\nn\\
 & &h_L\in ({SU(N)_L})_{local}~~~~~
h_R\in ({SU(N)_R})_{local}
\eea
In this way we have:
\be
U^\prime = g_L U g_R^\dagger
\ee
that is,
\be
U\in (\spin,\spin,0,0)
\ee
and, therefore, $U$ does not see the local symmetry (hidden gauge symmetry).
The Lagrangian (2.1) is obviously invariant under the discrete
transformation $U\to U^\dagger$, which corresponds to
(parity transformation):
\be
L\to R~~~~~
M\to M^\dagger~~~~~
R\to L
\ee

Proceeding in a completely standard way, we can build up
covariant derivatives with respect to the local group:
\def \LL{{{\bf L}_\mu}}
\def \RR{{{\bf R}_\mu}}
\bea
& &D_\mu L =\partial_\mu L - L\LL\nn\\
& &D_\mu R =\partial_\mu R - R\RR\nn\\
& &D_\mu M =\partial_\mu M - M\LL + \RR M
\eea
where $\LL$ and $\RR$ are the Lie algebra valued gauge fields of
$({SU(N)_L})_{local}$ and $({SU(N)_R})_{local}$ respectively.

We can now construct the invariants of our original group extended by the
parity operation defined in eq. (2.9). We find
\bea
{I}_1&=&\Tr ({L}^\dagger D_\mu {L}
       -M^\dagger D_\mu M-M^\dagger {R}^\dagger
          (D_\mu  {R}) M)^2\\
{I}_2&=&\Tr ({L}^\dagger D_\mu {L}+M^\dagger
           {R}^\dagger (D_\mu {R}) M)^2\\
{I}_3&=&\Tr ({L}^\dagger D_\mu {L}-M^\dagger {R}
         ^\dagger (D_\mu {R}) M)^2\\
{I}_4&=&\Tr (M^\dagger D_\mu M)^2
\eea

Using these invariants we can write the most general Lagrangian with at
most two derivatives in the form:
\be
{\cal L}=-\frac{v^2}{16} (a {I}_1+b {I}_2+c {I}_3
+d {I}_4)+~kinetic~terms~for~the~gauge~
               fields
\ee
where $a,~b,~c,~d$ are free parameters and furthermore the gauge coupling
constant for the fields $\LL$ and $\RR$ is the same.

It is not difficult to see that this Lagrangian is the same
one would obtain from the hidden gauge symmetry approach
\cite {bando}. The requirement of getting back
the non-linear $\sigma$-model in
the limit in which the gauge fields $\LL$ and $\RR$ are decoupled is
satisfied by imposing the following
relation among the parameters $a,b,c,d$
\be
a+ \frac{cd}{c+d}=1
\ee

We can now gauge the previous effective Lagrangian with respect to the standard
gauge group
$SU(3)_c\otimes SU(2)_L\otimes U(1)_R$ by the following substitutions:
\bea
D_\mu {L} &\to &  D_\mu {L} =
\dmu {L} -{L}
            (V_\mu-A_\mu)+{\bf A}_\mu {L}\\
D_\mu {R} &\to & D_\mu {R} = \dmu {R} -{R}
           (V_\mu+A_\mu)+{\bf B}_\mu {R}\\
D_\mu M &\to & D_\mu {M} =\dmu M -M (V_\mu-A_\mu)+(V_\mu+A_\mu) M
\eea
where $V_\mu=(\RR+\LL)/2$ and $A_\mu=(\RR-\LL)/2$ are the fields describing the
new vector and axial-vector resonances, whereas ${\bf A}_\mu$ and ${\bf B}_\mu$
are linear combinations of the gauge fields of the standard gauge group (see
later for a more precise definition).

The $SU(N)$ generators satisfy the algebra
\be
[T^A,T^B]=i f^{ABC}T^C
\ee
and are normalized by $\Tr(T^A T^B ) =\frac{1}{2}\delta^{AB}$.

The $V$ and $A$ bosons can be decoupled by sending $\gs\to\infty$. In this
limit, the mass of the $W$ bosons is the SM mass with
$v\simeq 246~GeV$.

 The effective Lagrangian (2.15)
describes the interactions among the SM gauge
bosons, the new vector and axial-vector resonances and the Goldstone bosons.

We will work in the unitary gauge both for the standard gauge fields, defined
by $\pi^a=0$, and
for the $V$ and $A$ bosons given by the following choice:
\bea
{L}&=&{R}^\dagger=\exp\left({\frac{2 i \pi^A T^A}{v}}
{\frac {d}{c+d}}
\right)\\
M &=& \exp\left(-{\frac{4 i \pi^A T^A}{v}}
{\frac {c}{c+d}}
\right)
\eea

To avoid cumbersome notations we we shall restrict in the following to
$N=8$, to which the quantitative discussion will be limited, except for
some occasional more general remarks.

We will denote the $SU(8)$ gauge fields as
$V^A=(V^a,{\tilde V}^a,V_D,V_8^\alpha,V_8^{a \alpha},V_3^{\mu i},
{\bar V}_3^{\mu i})$, where $\mu=(0,a)$ ($a$ being an $SU(2)$ index),
 and $i=1,2,3$ is a color index. An analogous notation will be used for
 $A^A$ and  the Goldstone bosons $\pi^A$. The $SU(8)$ generators are shown
for convenience in Appendix A.
In the following we will make use of the notations:
\begin{eqnarray}
{\bf A}_\mu &=&
2 i g W^a_\mu T^a+i \sqrt{2} g_s G^\alpha_\mu T^\alpha_8+2 i y\gp
Y_\mu T_D\\
{\bf B}_\mu &=&
2 i \gp Y_\mu(T^3 + y T_D)+
i  \sqrt{2} g_s G^\alpha_\mu T^\alpha_8\\
V_\mu &=& i \gs V_\mu^A T^A \\
A_\mu &=& i\gs A_\mu^A T^A
\end{eqnarray}
with  $A,B=1,\ldots ,63$; $a,b=1,2,3$; $\alpha ,\beta =1,\ldots ,8$ and
$y =1/\sqrt{3}$ (for future convenience,
we have not substitute its numerical value).
We have denoted by $W,~Y,~G$ the standard model (SM)
gauge bosons and by $g,~\gp,g_s$ their coupling constants,
 while $\gs$ is the self coupling of the $V$ and $A$ bosons.
Let us consider the quadratic terms in eq. (2.15) to see
which is the structure of the mixing between the SM gauge bosons and
the new vector and axial-vector resonances:
\bea
{\cal L}^{(2)} &=&
\frac{v^2}{8}\Big[a(g W^a-\gp \delta^{a3} Y)^2
   +b(g W^a+\gp \delta^{a3} Y-\gs V^a)^2+b (2 y \gp Y-\gs V_D)^2\nn\\
& &+b (\sqrt{2} g_s G^\alpha-\gs V^\alpha_8)^2+b \gs^2
\sum_{A\ne a,D,\alpha}   (V^A)^2+
c(g W^a-\gp\delta^{a3} Y+\gs A^a)^2\nn\\
& &+c \gs^2
\sum_{A\ne a}   (A^A)^2+
d\gs^2 (A^A)^2 \Big]+~kinetic~terms
\eea

In the charged sector the mixing terms are the same as in
ref. ${\cite {assiali}}$.

In the neutral sector the mixing involves the fields
$W^3,~Y,~V^3,~V_D,~A^3$.
The mixing with $V_D$, which makes the gauge boson sector of
the $SU(8)$-BESS different in a non trivial way from the model
based on $SU(2)_L\otimes SU(2)_R$, is parametrized by $y$.

The mass eigenvalues in the limit $\gs\rightarrow\infty$ are (see Appendix B):
\bea
M_{\gamma}^2 &=& 0\\
M_{W}^2&\simeq&\frac{v^2}{4}g^2\left (1-(\ggs)^2 \left(\frac{1}
{1-r_V}-\frac{z^2}
{1-r_A}\right )\right)\\
M_Z^2 &\simeq& \frac{v^2}{4} G^2 \left( 1-(\ggs)^2 \left(\frac{1}
{c^2_\theta-r_V} (c^2_{2\theta}+4 y^2 s_\theta^4)+ \frac{z^2}
{c^2_\theta-r_A} \right)\right)\\
M_V^2&\simeq&\frac{v^2}{4}\frac{g^2}{r_V}\\
M_A^2&\simeq&\frac{v^2}{4}\frac{g^2}{r_A}
\eea
where $G=\sqrt{g^2+\gp^2}$, $s_\theta=\sin\theta=\gp/G$, $x=g/\gs$, and
\be
r_V=\frac{x^2}{b}~~~~~r_A=\frac{x^2}{c+d}~~~~~z=\frac{c}{c+d}
\ee
Notice that, in the limit considered,
all vector and all axial-vector masses
are degenerate.

Finally we observe that the mixing angle for the colored sector
in the large $\gs$ limit is  $\simeq -\sqrt{2} g_s/\gs$.
The linear combinations corresponding to the gluons remain massless whereas
the orthogonal ones are degenerate in mass with all the other $V$ bosons (in
the large $\gs$ limit).

For simplicity we do not add a direct coupling of the $V$ and $A$ bosons
to the fermions.
Therefore only the gauge bosons $V^a,~V_D,~V_8^\alpha,~A^a$
can be produced via quark-antiquark, through their mixing with the SM
gauge bosons.

For what concerns the low energy interactions, in the charged sector things go
 exactly as in the $SU(2)$-BESS model
\cite{assiali}, and the charged current-current interaction coincides with the
SM one by the identification $\sqrt{2} G_F=1/v^2$.

The neutral current-current interaction strength is given by:
\be
\frac{4}{\sqrt{2}}\rho~ G_F=
\frac{2}{v^2}({\bar{\cal M}}^{-1})_{11}
\ee
with ${\bar{\cal M}}$  given in Appendix B.
Also in this case (as in ref. \cite{assiali})
 we get $({\bar{\cal M}}^{-1})_{11}=1$
and therefore $\rho=1$. This is due to the global unbroken diagonal $SU(8)$.
The same is true in general for $SU(N)\otimes SU(N)$.
Finally, from the neutral current, one can extract the expression
for the electric charge
\be
e=\gp c_\theta c_\psi c_\eta
\ee
with the angles $\psi$ and $\eta$
defined in Appendix B. The difference with respect to the
$SU(2)$-BESS is contained in the mixing angle $\eta$ (which vanishes
for $y=0$) corresponding to the $V_D$ contribution.

\resection{Gauge boson self-energies}

\def\bm{{\bar M}}
\def\BV{{\bf B}}
\def\WV{{\bf W}}
\def\VV{{\bf V}}
\def\AV{{\bf A}}
We will now compute the corrections to the self-energies of the SM
gauge bosons from the new vector and axial-vector resonances and from the
pseudogoldstones.
We will consider first the tree-level contribution induced by the mixing
of the $V$ and $A$ bosons with the $W$, $Z$ and $\gamma$.

We define the scalar part of the vector boson self-energies through the
relation:
\be
\Pi_{ij}^{\mu\nu}(q^2)=-i\Pi_{ij}(q^2)g^{\mu\nu}+q^\mu q^\nu~terms
\ee
where the indices $i$ and $j$ run over the ordinary gauge vector bosons.
The self-energy corrections are obtained by
separating out the SM terms from the
bilinear part of the Lagrangian, given in eq. (2.27),
and by computing, with the remaining terms, all the tree-level self-energy
graphs which are one-particle-irreducible with respect to the lines $W$,
$\gamma$ and $Z$. This is equivalent to solve the equations of motion
for the fields $V^\pm$, $A^\pm$, $V^3$, $V_D$ and $A^3$ by using again
the quadratic Lagrangian.

Since the mixing in the charged sector of $SU(8)$-BESS is the same as in
$SU(2)$-BESS,
we get the same result for the $WW$ self-energy correction as in ref.
\cite{self}:
\be
\Pi_{WW}(q^2)=-x^2 \left( M^2_V \frac{q^2}{q^2- M^2_V}+
                      z^2 M^2_A \frac{q^2}{q^2- M^2_A}\right)
\ee

In the neutral sector we choose to work with the SM combinations:
\bea
& &W^3 =c_\theta~ Z+s_\theta~ \gamma\\
& & Y =-s_\theta~ Z+c_\theta~ \gamma
\eea

The solutions of the equations of motion for the new bosons are:
\bea
V^3&=& x\frac{ M^2_V}{M^2_V-q^2} (W^3+ \tan\theta ~Y)\\
V_D&=&2x y \tan\theta \frac{ M^2_V}{M^2_V-q^2}Y\\
A^3&=& -x z \frac{ M^2_A}{M^2_A-q^2} (W^3-  \tan\theta  ~Y)
\eea
By substituting these relations in the equations of motion for the
$W^3$ and $Y$ fields we can read the expression for the self-energy
corrections $\Pi_{33}$, $\Pi_{30}$ and $\Pi_{00}$ as
\bea
\Pi_{33}(q^2)&=&\Pi_{WW}(q^2)\\
\Pi_{30}(q^2)&=&-\frac{ s_\theta}{c_\theta}{\bar M}^2_W x^2
\left( \frac{1}{r_V}
\frac {q^2}{q^2- M^2_V}-\frac{z^2}{r_A}
\frac {q^2}{q^2- M^2_A}\right)\\
\Pi_{00}(q^2)&=&-\frac{ s^2_\theta}{c^2_\theta}{\bar M}^2_W x^2
\left((1+4y^2)\frac{1}{r_V}\frac {q^2}{q^2- M^2_V}+\frac{z^2}{r_A}
\frac {q^2}{q^2- M^2_A}\right)
\eea
where ${\bar M}^2_W=(v^2 g^2)/4$.

It is convenient to introduce the following combinations
(see ref. \cite{altarelli}):
\bea
\epsilon_1&=&\frac{\Pi_{33}(0)-\Pi_{WW}(0)}{M^2_W}\\
\epsilon_2&=&\frac{\Pi_{WW}(M^2_W)-\Pi_{WW}(0)}{M^2_W}-
\frac{\Pi_{33}(M^2_Z)-\Pi_{33}(0)}{M^2_Z}\\
\epsilon_3 &=&\frac {c_\theta}{s_\theta}
\frac{\Pi_{30}(M^2_Z)-\Pi_{30}(0)}{M^2_Z}
\eea
The new $V_D$ boson does not affect the $\epsilon_i$ parameters because
it contributes only to $\Pi_{00}$. Therefore
we recover the same expressions as for $SU(2)$-BESS, or
\be
\epsilon_1=0~~~~\epsilon_2\simeq 0,~~~~\epsilon_3\simeq x^2 (1-z^2)
\ee
where the last two results are obtained for large $M_{V,A}$
($M_{V,A}>>M_{W,Z}$).

The additional contribution to $\Pi_{00}$ affects the electric
charge  and the $Z$ mass definition, as can be seen by eqs. (2.35) and (2.30).

The reason we get the same result as in ref. \cite{self} is that
the new $V_D$ boson mixes only with $Y$ (see eq. (2.27)).
This means that our result can be extended to
a $SU(N)_L\otimes SU(N)_R$ model. This is due to the fact
that in these models the new fields associated to the
diagonal generators will be mixed only with the hypercharge field $Y$.

It is interesting to notice that the same result for $\epsilon_3$ can be
obtained by using the dispersive representation which had been already given
in refs. \cite{peskin} \cite{Cahn-Suzuki}.
In fact, by using the relation $\epsilon_3=(\alpha_{em}/4 s^2_\theta)S$
we get:
\be
\epsilon_3=
-\frac{g^2}{4\pi}\int_0^\infty \frac{ds}{s^2}\left[ {\rm Im}~ \Pi_{VV}(s)-
{\rm Im}~ \Pi_{AA}(s)
\right]
\ee
where $\Pi_{VV(AA)}$ is the correlator of the vector (axial-vector) currents.
If we assume vector meson dominance, we can saturate the eq. (3.15)
with the $V$ and $A$ resonances:
\be
{\rm Im}~\Pi_{VV (AA)}(s)=-\pi g^2_{V(A)}\delta(s-M^2_{V(A)})
\ee
where $g_{V}$ and $g_A$ are the couplings of the
vector and axial-vector currents to the $V$ and $A$ fields respectively.
These couplings can
be obtained directly from eq. (2.15), using eqs. (2.31-32):
\be
g_V=-2\frac{M^2_V}{\gs}~~~~~~~~g_A=2 z \frac{M^2_A}{\gs}
\ee
and substituting eqs. (3.16-17) in eq. (3.15)
 we get the tree-level contribution to
$\epsilon_3$ given in eq. (3.14).

We will now consider the one-loop contributions to the self-energies in order
to calculate the corrections to the $\epsilon_i$.

There are four kinds of loops. The first is a loop of Goldstones, the second
is a loop of one Goldstone boson and one $V$ boson, the third one
is a loop of one Goldstone boson and one $A$ boson and the fourth is a
tadpole of Goldstones.
Also, since we will work with
the fields appearing in the Lagrangian of eq. (2.15) which
are not the mass eigenstates,  we will take into account
all the possible mixings on the external legs.

For the calculation of the loop contributions to $\epsilon_i$ it is
useful to compute from the effective Lagrangian given in eq. (2.15) (in
the unitary gauge given by eqs. (2.25-26) and with $\pi^a=0$)
 trilinear and quadrilinear
terms in the colorless sector.

The trilinear terms in the gauge boson neutral sector are ($y=1/\sqrt 3$) :
\bea
i{\cal L}_{neutr}^{(3)} & = &\Big\{ \Big[ g_{W\pi\pi}(W^3_\mu+\tan\theta~Y_\mu)
+g_{V\pi\pi}V^3_\mu\Big](-{\tilde\pi}^-\dmus{\tilde\pi}^+
-\pi_8^{\alpha -}\dmus\pi_8^{\alpha +}\nn\\
& &+{\bar P}_3^{-i}\dmus P_3^{-i}-{\bar P}_3^{+i}\dmus P_3^{+i})
+(-\frac{4}{3}g_{W\pi\pi}\tan\theta~ Y_\mu-\frac{2}{\sqrt 3}g_{V\pi\pi}
V_{D\mu})
\nn\\
& & ({\bar P}_3^{0i}\dmus  P_3^{0i}+{\bar P}_3^{3i}\dmus { P}_3^{3i}+
 {\bar P}_3^{-i}\dmus P_3^{-i}+{\bar P}_3^{+i}\dmus P_3^{+i})\nn\\
& &+\Big[g_{WV\pi}( W^3 -\tan\theta ~Y)-g_{VA\pi}A^3\Big]
({\tilde V}^+ {\tilde\pi}^- +
     V_8^{\alpha +} \pi_8^{\alpha -}-{\bar P}_3^{-i}{\cal V}_3^{-i}+
      {\bar P}_3^{+i}{\cal V}_3^{+i})\nn\\
& &+\Big[-g_{WA\pi}( W^3 +\tan\theta ~Y)+g_{VA\pi}V^3\Big]
({\tilde A}^+ {\tilde\pi}^- +
     A_8^{\alpha +} \pi_8^{\alpha -}-{\bar P}_3^{-i}{\cal A}_3^{-i}+
      {\bar P}_3^{+i}{\cal A}_3^{+i})\nn\\
& &+\Big[-\frac{4}{3}g_{WA\pi}\tan\theta ~Y+\frac{2}{\sqrt 3}g_{VA\pi}V_D\Big]
({\bar P}_3^{0i}{\cal A}_3^{0i}+{\bar P}_3^{3i}{\cal A}_3^{3i}+
           {\bar P}_3^{-i}{\cal A}_3^{-i}+{\bar P}_3^{+i}{\cal A}_3^{+i})\nn\\
& &-h.c.\Big\}\nn\\
\eea

The notation used here for the fields is given in Appendix C, in particular
${\cal V}_3^{\mu i}$ and
${\cal A}_3^{\mu i}$ are  defined analogously as
 $P_3^{\mu i}$ and
\be
g_{W\pi\pi}=\frac{g}{2}\left[1-\frac{1}{2}\frac{x^2}{r_V}(1-z^2)\right]~~~~
g_{V\pi\pi}=\frac{\gs}{4}\frac{x^2}{r_V}\left(1-z^2\right)\ee
\be
g_{WV\pi}=\frac{v}{4}g\gs\frac{x^2}{r_V}(1-z^2\frac{r_V}{r_A})~~~
g_{VA\pi}=\frac{v}{4}z\gs^2\frac{x^2}{r_V}\left( \frac{r_V}{r_A}-1\right)~~~
g_{WA\pi}=x~ g_{VA\pi}
\ee
We have also used the relations
\be
a=1-\frac{x^2}{r_A}(1-z)z~~~~~~b=\frac{x^2}{r_V}~~~~~~c=z\frac{x^2}
{r_A}~~~~~~d=\frac{x^2}{r_A} (1-z)
\ee
obtained from eqs. (2.16) and (2.33).
Notice that
we have not written down the trilinear terms coming from the kinetic terms.

The relevant quadrilinear couplings turn out to be
those involving two colorless
gauge fields
and two Goldstone
bosons.
Furthermore, only the terms containing the
$V^3$, $V^\pm$, $V_D$, $A^3$, $A^\pm$ fields
contribute to the calculation of the correction to the $W, Y$ self-energies.
In fact, only these components of $V^A$ and
$A^A$ mix with the standard gauge bosons.
We list here
the terms which are relevant for the computation of $\epsilon_i$:

\bea
{\cal L}^{(4)} & = &\Big\{ \Big[ g_{VV\pi\pi}(V_3^2-A_3^2)
+W_3(g_{WV\pi\pi}V_3-g_{WA\pi\pi}A_3)\Big]\nn\\
& &\times({\tilde\pi}^-{\tilde\pi}^+
+\pi_8^{\alpha -}\pi_8^{\alpha +}
+{\bar P}_3^{+i} P_3^{+i}+{\bar P}_3^{-i}P_3^{-i})\nn\\
& &+\Big[g_{VV\pi\pi}(V^+V^--A^+A^-)+W^+(g_{WV\pi\pi}V^--
g_{WA\pi\pi}A^-)\Big]\nn\\
& &\times({\tilde\pi}^-{\tilde\pi}^+
+({\tilde\pi}^3)^2+\pi_8^{\alpha -}\pi_8^{\alpha +}
+(\pi_8^{\alpha 3})^2\nn\\
& &+{\bar P}_3^{+i} P_3^{+i}+{\bar P}_3^{-i}P_3^{-i}+
2 {\bar P}_3^{3i}P_3^{3i})+h.c.\Big\}
\eea
where
\be
g_{VV\pi\pi}=\frac{\gs^2}{8}z^2\frac{x^2}{r_V}
\left(\frac{r_V}{r_A}-1\right)
\ee
\be
g_{WV\pi\pi}=\frac{g\gs}{8}\frac{x^2}{r_V}
\left(1+z^2-2z^2\frac{r_V}{r_A}
\right)
\ee
\be
g_{WA\pi\pi}=\frac{g\gs}{8}z \frac{x^2}{r_V}\left((1+z^2)\frac{r_V}{r_A}-2
\right)
\ee
Notice that
we have not written down the quadrilinears coming from the kinetic terms.
Actually, as we will see, they do not contribute to the calculation of
$\epsilon_i$.

We have calculated the loop integrals by using a cut-off $\Lambda$, keeping
also the finite terms,
assuming a non vanishing mass for the Goldstone bosons and in the limit
$M^2_{V,A}>>q^2$.

\resection{One-loop contribution to $\epsilon_3$}

In Figs. 1-2  we show the graphs contributing to $\epsilon_3$ at one-loop
level.

We have separately drawn in Fig. 2 the diagrams depending on
the $V$ and $A$ self-energies. In order to
correctly evaluate their contributions to $\epsilon_3$ one has to properly
renormalize the BESS model at one-loop and then extract the finite
terms coming from these graphs which are not absorbed in the
redefinition of the BESS parameters.
In the present calculation of $\epsilon_3$
we have not
used this procedure
but we have simply tried to estimate the dominant contribution of the
$V$ and $A$ self-energy graphs by using
a dispersive representation for $\epsilon_3$.
The calculation is given in Appendix D and further comments will be given in
Sect. 6.

We have not listed the tadpole loops of pseudo-Goldstone bosons (PGB) and
of vector and axial-vector bosons. In fact,
due to the form of the $\epsilon_3$ given in eq. (3.13),
the seagulls do not contribute because they are independent of $q^2$
in the $M^2_{V,A}>>q^2$ limit.

We give here the result for the one-loop contribution obtained by
summing  up the graphs of Fig. 1.
We have regularized the integrals with a cut-off $\Lambda$
and we have considered a non-vanishing mass $m_\Pi$ for the PGB's.
In fact, since $\epsilon_3$ is an isospin symmetric observable, we can
consider the same mass for all the PGB's. Here is the result:

\bea
\epsilon_3^{loop} &\simeq &
\frac{g^2}{16\pi^2} \frac{5}{8}\Big \{\left
(\log\frac{\Lambda^2}{m_\Pi^2}-\gamma\right)
\Big[2-\frac{1}{2} \frac{x^4}{r_V^2}(1-z^2)^2\nn\\
& &-\frac{x^2}{r_V}(1-z^2\frac{r_V}{r_A})
 \Big(1 -z^2 -z^2(1-\frac{r_V}{r_A})\Big)-z^2
\frac{x^2}{r_A}(1-\frac{r_A}{r_V})^2\Big]\nn\\
& &-\frac{x^2}{r_V}(1-z^2\frac{r_V}{r_A})
\Big(1 -z^2 -z^2(1-\frac{r_V}{r_A})\Big)\Big(A(m_\Pi^2,M_V^2)+1\Big)\nn\\
& &-z^2\frac{x^2}{r_A}(1-\frac{r_A}{r_V})^2\Big(A(m_\Pi^2,M_A^2)+1\Big)
\Big\}
\eea
where $\gamma$ is the Euler's constant $(\simeq 0.577)$, and
\bea
A(m_\Pi^2,M^2)&=&
\frac{M^6+9  M^4 m_\Pi^2}{(M^2-m_\Pi^2)^3}
\log{\frac{m_\Pi^2}{M^2}}\nn\\
& &+\frac{1}{6 (M^2-m_\Pi^2)^3} (m_\Pi^6-27 m_\Pi^4 M^2-9 m_\Pi^2 M^4
+35 M^6)
\eea

Notice that, by decoupling the vector and axial-vector resonances (that is
by taking $x=0$ in eq. (4.1)),
one recovers the $SU(8)$ technicolor result (see for
instance ref. \cite{Golden-Randall}). The same result is obtained in the
completely degenerate case: $z=1$ and $r_V=r_A$. In fact the vector and the
axial-vector contributions cancel and only the PGB loops remain.
Also, for $z=0$ the axial-vector resonances decouple.

\resection{Effects of the isospin violating terms: one-loop contribution to
         $\epsilon_1$}

For the calculation of $\epsilon_{1}^{loop}$ one needs $\Pi_{33}$ and
$\Pi_{WW}$ at the one-loop level.

The graphs contributing to $\Pi_{33}$ are those given in Figs. 1-2
 with the substitution
($Y \to W^3$)
on the external leg, plus the tadpole graphs of Fig. 3.
Also, as can be seen from the Lagrangian terms in eqs. (3.18) and (3.23),
the graphs contributing to $\Pi_{WW}$ and to $\Pi_{33}$
are of the same kind.
So, if one assumes
mass degeneration
for the $SU(2)$ multiplets of the pseudo-Goldstone bosons, it is easy to
show that one gets again
\be
\Pi_{WW}^{loop}=\Pi_{33}^{loop}
\ee
and so $\epsilon_1^{loop}=0$.

In the case in which the PGB masses are not $SU(2)$
symmetric one expects non vanishing
contributions to $\epsilon_1^{loop}$  coming from the splitting.

An important property
of the $SU(2)$ triplets and of the associated singlets of PGB's is the validity
of sum rules for their masses
following from the symmetry structure of the theory.
For a generic quadruplet $P^+,P^-,P^0,P^3$ we can
show that \cite{masse}
\be
\mpp^2+\mpm^2=\mpz^2+\mpt^2
\ee

In fact in the chiral limit one has a
$G=SU(2)_L\otimes SU(2)_R$ symmetry for each fermionic doublet.
Then the mass matrix must be a combination of a scalar and of
generators of $G\sim O(4)$ which commutes with the electric charge.
Therefore the general structure
of the mass matrix is
\be
M^2=A +BK_3+CT_3
\ee
with  $T_3=(1/2)(T_{3L}+T_{3R})$ and $K_3=(1/2)(T_{3L}-T_{3R})$.
Notice that the $({1\over 2},{1\over 2})$ representation of
$G$ to which the Goldstone
bosons  belong, decomposes as
${\underline 1}+{\underline 3}$ with respect to $O(3)\subset
O(4)$ and therefore
the following sum rules for the masses are implied
for all the quadruplets:
\be
\sum m^2_\Pi(T_3=0)=\sum m^2_\Pi(K_3=0)
\ee
which is equivalent to eq. (5.2) (see also \cite{renken}).

An important consequence of the validity of the previous sum rules is that
the dependence on the cut-off $\Lambda$ (introduced to regularize the one-loop
integrals) cancels and we get an ultraviolet finite result (see also
\cite{renken}).
Adding all the contributions from the different loops
we get (for a single quadruplet):
\def\dmz{\delta_0}
\def\dmt{\delta_3}
\bea
\epsilon_1^{loop}&\simeq &\frac{g^2}{256\pi^2}\frac{\mpp^2}{{\bar M}^2_W}
 \Big(-\dmz^+\log\frac{\mpp^2}{\mpm^2}+ F(P^+,P^0)+F(P^+,P^3)\nn\\
& &+\frac{x^2}{r_V}(1-z^2)^2\Big[\dmz^+\log\frac{\mpp^2}{\mpm^2}
+(1+\dmz^+)\log(1+\dmz^+)\nn\\
& &+(1+\dmt^+)\log(1+\dmt^+)
 +2~g(\eta^+)-(1+\dmt^+)~g(\eta^3)\nn\\
& &-(1+\dmz^+)~g(\eta^0)\Big]\Big)
+\Big(P^+\longleftrightarrow P^-\Big)
\eea
where
\be
F(P^+,P^0)=1-
\frac{(1+\dmz^+)^2}{\dmz^+}\log(1+\dmz^+)
\ee
and
\be
g(\eta)=\frac{1}{1-\eta}(1-\frac{\eta}{4})\log \eta~~~~~~
\eta=\frac{m^2_P}{M^2_V}
\ee
with
\be
\dmz^+=\frac{\mpz^2-\mpp^2}{\mpp^2}~~~~~~~\dmt^+=\frac{\mpt^2-\mpp^2}{\mpp^2}
\ee
and $\dmz^-$ and $\dmt^-$ analogously defined.
The total result is obtained by
collecting the similar contributions from all the quadruplets
and taking into account the appropriate multiplicity in the color sector:
\be
\epsilon_1=\epsilon_1(col.~singl.)+8\epsilon_1(col.~oct.)+
3\epsilon_1(col.~tripl.)+3\epsilon_1(col.~antitripl.)
\ee

Notice that the one-loop result does not depend on the mass of the
axial-vector resonances. This is due to cancellation among the terms coming
{}from the loops of $A$ bosons and PGB's. The presence of
the axial-vectors is signalled only by the $z$ parameter
entering into the quadrilinear
couplings (see eqs. (3.24-25)). For $z=1$, again, the new gauge boson
contributions cancel and we get the result from the PGB loops.

We have also calculated the one-loop contribution to $\epsilon_2$.
As it is clear by comparing the definitions in eq. (3.12), we have
that, within the BESS model,
 $\epsilon_2$ is depressed with respect to
$\epsilon_1$ by a factor $\sim~M^2_W/M^2_V$. Therefore the further
restrictions coming from the numerical analysis based on $\epsilon_2$
are quite irrelevant and we will not consider them.

In ref. \cite{masse} we have developed a framework for a quantitative
evaluation of the PGB masses, including, besides the gauge contribution
\cite{massepeskin}, also the contribution from those interactions which are
responsible for the masses of the ordinary fermions.
In this way, all those states which, neglecting
such Yukawa terms, would remain massless, tend to acquire  mass terms which
are close to those of the heaviest fermions of the theory. In such a scheme
a splitting among the PGB masses is unavoidable
since it is due to the mass difference between the top and the bottom quarks.
Furthermore in this case, mixings between the neutral components
of the $SU(2)$ triplets and the associated singlets appear.

\resection{Numerical results}

We now give a quantitative estimate of the parameters $\epsilon_3$
and $\epsilon_1$ as predicted by the BESS model.
The BESS parameters are
$M_V$, $M_A$, $\gs$ and $z$ and one must also add to them the
values of the pseudo-Goldstone masses and the cut-off $\Lambda$.
To reduce the BESS parameter space
we will assume the validity of the Weinberg Sum Rules (WSR),
which have been shown to hold in asymptotically free gauge theories
\cite{losecco}:
\bea
& &\frac{1}{\pi}\int_0^\infty \frac{ds}{s}\left[ {\rm Im}~ \Pi_{VV}(s)-
{\rm Im}~ \Pi_{AA}(s)
\right]=v^2\nn\\
& &\frac{1}{\pi}\int_0^\infty ds \left[ {\rm Im}~ \Pi_{VV}(s)-{\rm Im}~
 \Pi_{AA}(s)
\right]=0
\eea
where $\Pi_{VV(AA)}$ is the correlator of the vector (axial-vector) currents.

If we also assume vector meson dominance, we can saturate the eqs. (6.1)
with the $V$ and $A$ resonances by using the relation (3.16).
In this way we get
\be
M^2_A=\frac{M^2_V}{z}~~~~~~~~\gs=2\frac{ M_V}{v} \sqrt{1-z}
\ee
with $0<z<1$.

With a further specialization of the BESS parameters, one can reproduce
a standard technicolor scheme \cite{techni}.
In the case of a $SU(N_{TC})$
scaled up version of QCD, one has:
\bea
M^2_{\rho TC}&=&M^2_0 \frac{3}{N_{TC}}\frac{4}{N_d}\nn\\
M^2_{A_{1}TC}&=&2 M^2_{\rho TC}
\eea
where $N_d$ is the number of technidoublets (in the present calculation
$N_d=4$) and $M_0$
is a scale parameter of the order of $1~TeV$.
By comparing with the eq. (6.2) one gets
\be
z=\frac{1}{2}~~~~~~~~
\frac{g}{\gs}=\frac{{\bar M}_W}{M_0}\sqrt{\frac{N_{TC}N_d}{6}}
\ee

Our strategy will be to work in the context of validity of the WSR's leaving
$M_V$ and $z$ as free parameters.
Therefore we remain with a two-dimensional parameter space
plus the cut-off and the PGB masses.

Before comparing the prediction of this model for the observables $\epsilon_i$,
we must consider the radiative corrections coming from the standard model
contributions.

The BESS model has no elementary scalars and it is not a renormalizable
theory. Radiative corrections for BESS can be defined only if one considers
this model as a cut-off theory. We will assume the same one-loop
radiative corrections as in the SM by interpreting the Higgs mass $m_H$ as
the cut-off $\Lambda$ used to regularize the theory.
In particular, in our numerical estimations, we have used the
SM radiative corrections to the $\epsilon_i$ as given in the last of
ref. \cite{altarelli}.

The experimental limits on the $\epsilon_i$ come from the measured values
of some weak interaction observables. The minimal set is
given by $M_W/M_Z,~ A_{FB}^l,~\Gamma_l$. One can also include
the $A^\tau_{pol.},~A_{FB}^b$ and the low energy data, in particular
neutrino-nucleus deep inelastic scattering and parity violation in Cs atoms.
In the following analysis we will use the values obtained considering
this larger set \cite{moriond}:
\be
\epsilon_1= (0.16\pm 0.32) 10^{-2}~~~~
\epsilon_2= (-0.72\pm 0.79) 10^{-2}~~~~
\epsilon_3= (0.00\pm 0.43) 10^{-2}
\ee

Let us first consider the bounds on the BESS parameters coming from the
estimation of the corrections to $\epsilon_3$.

In the following numerical analysis we will neglect the contribution
coming from the graphs of Fig. 2. The reason is that the estimate we have
given in Appendix D is reliable only when $M_V$ and $M_A$ are above threshold
for the decay in $\pi\pi$ and in $\pi V$ respectively.
This implies that the representations given in eqs. (D.1) and (D.4)
cannot be valid for generic values of $m_\Pi$ and $z$.
However we can make the following comments.
It has already been noticed \cite{Cahn-Suzuki}
 that, if the
PGB masses are large, $\delta\epsilon_3^V$ can be negative
and decreases the tree level contribution by a sizeable amount. For
instance, taking $M_V=1200~GeV$, $z=0.5$, and $m_\Pi=300~GeV$, one finds
$\delta\epsilon_3^V=-0.002$, to be compared to $(\epsilon_3^V)_{tree}=0.009$.
In this example the conclusion holds true even by including the
axial-vector contribution. One has $\delta\epsilon_3=-0.001$ and
$(\epsilon_3)_{tree}=0.0067$. More generally, by studying numerically
$\delta\epsilon_3$ in eqs. (D.1)-(D.7), one sees that, for fixed $M_V$
and $z$, there is a window for $m_\Pi$, usually requiring $m_\Pi$ to be
sufficiently large, where $\delta\epsilon_3$ is negative. In the cases
we have analyzed, the tree-level result might be lowered by about 15\%.
Positive contributions to $\epsilon_3$ could also arise for small
PGB masses $m_\Pi$.

Starting from eq. (4.1) and substituting the relations (6.2) which follow
{}from the WSR's,
 we obtain an expression for $\epsilon_3^{loop}$ which depends on
the cut-off $\Lambda$, the  mass of the PBG's $m_\Pi$
(which we assume degenerate),
 the mass of the $V$ bosons and the parameter $z$. Actually
the dependence on $\Lambda$ and on $M_V$ is very weak (in the range of values
of interest that is $\sim 1-2~TeV$) and also the dependence on $m_\Pi$ is not
too strong.

In Fig. 4 we plot the one-loop contribution (dashed line)
 to $\epsilon_3$ as a
function of $z$ for $\Lambda=1.5~TeV$, $M_V=1~TeV$ and $m_\Pi/\Lambda=0.35$.
  Also reported in Fig. 4 is the tree-level contribution as given
in eq. (3.14) with the substitution (6.2) summed with
 the SM radiative corrections
(dotted line). Finally the total correction to $\epsilon_3$ at
one-loop level  is given
by the solid line.
By comparing this result with the experimental data given in eq. (6.5)
we see how the one-loop contribution, which turns out to be negative,
 leads to a better agreement
(in the figure the experimental $1\sigma$ band is shown).

This can be seen also in Fig. 5
 where the lower bound on $M_V$ as a function of $z$
coming from  the measure of $\epsilon_3$ is plotted.
Again the dotted line is the lower bound coming from the tree-level
plus the SM radiative corrections.
 The inclusion of the one-loop contribution enlarge the allowed region
 (the region is at 90$\%$ C.L.).
In the figure it is shown how the bound varies for different values
of $m_\Pi/\Lambda$, ranging from 0.10  (dashed line) to 0.35 (solid line).
We see that the dependence is very weak.
Here $\Lambda=1.5~TeV$.
Also shown
is the case of a standard
technicolor theory with one family of technifermions (black
dot in the figure) which is now marginally included in the
 90$\%$ C.L. region. In fact the
positive contribution of the PGB loops is overcompensated by the negative one
coming from the loops containing the vector and the axial-vector resonances.
Notice that the region shown in Fig. 5 corresponds to a pessimistic estimate
in view of the
previous comment on the $V-A$ self-energy contributions.

It is not difficult to see that each $SU(2)$ doublet of Pseudo Goldstone
bosons, vector and axial-vector resonances gives a negative
contribution to $\epsilon_3$. Therefore, in a model $SU(N)_L\otimes SU(N)_R$
the value of $\epsilon_3$ decreases by increasing $N$.

Let us quantitatively analyze the corrections to $\epsilon_1$ coming from
isospin violating terms.
In order to do this, one has to specialize the spectrum of the masses
of the  PGB's.

As stressed in Section 5, the PGB mass spectrum must satisfy the
sum rules (5.2).
A possible simplified parameterization
which is consistent with the sum rules and
the discussion made in ref. \cite{masse} is the following:
\bea
& &m^{2}({\tilde \pi}^{\pm})=\Delta m^2\\
& &m^{2}\left(\frac{{\tilde\pi}^{3}-\pi_{D}}{\sqrt{2}}\right)\simeq 0\\
& &m^{2}\left( \frac{{\tilde\pi}^{3}+\pi_{D}}{\sqrt{2}}\right)=2\Delta m^2\\
%& &m^{2}(\pi_{8}^{\alpha \pm})=
%m^{2}\left(\frac{\pi_8^\alpha+\pi_8^{\alpha 3}}{\sqrt{2}}\right)=
%m^{2}\left(\frac{\pi_8^\alpha-\pi_8^{\alpha 3}} {\sqrt{2}}\right)\\
& &m^{2}\left( P_{3}^{\pm i}\right)=
\frac{\Lambda^{2} g_s^2}{2 \pi^2} +\Delta m^2\\
& &m^{2}\left( \frac{P^{0i}_3-P_3^{3i}}{\sqrt{2}}\right)
=\frac{\Lambda^{2} g_s^2}{2 \pi^2} \\
& &m^{2}\left( \frac{P^{0i}_{3}+P_{3}^{3i}}{\sqrt{2}}\right)
=\frac{\Lambda^{2} g_s^2}{2 \pi^2} +2\Delta m^2
\eea
where we have used
the notation given in Appendix C.
In this example,
the masses of the colorless sector depend only on $\Delta m$.
In eq. (6.7), for numerical purposes, we have assumed
a zero mass because in ref. \cite{masse} we have found that there exists an
upper bound of 9 $GeV$ for $\Lambda=1~TeV$ (an analogous estimate comes also
{}from other mechanisms of mass generation, see for instance
ref. \cite{ellis}).

The colored states
get mass also from QCD corrections \cite{massepeskin}.
We have not included the octet states because for them the splitting is
negligible (they receive the strongest QCD correction)
and so they do not significatively contribute to $\epsilon_1$.
For the color triplets we have assumed the same splitting $\Delta m$
as for the colorless sector
while the QCD contribution is given in terms of
$g_s$, which is the $SU(3)$
gauge coupling constant (we will use $\alpha_s=0.12$),
and in terms of $\Lambda$ which is the ultraviolet cut-off introduced
to regularize the quadratic divergence in the one-loop effective potential
\cite{massepeskin} \cite{masse} (we will take $\Lambda=1~TeV$).

If one follows ref. \cite{masse} $\Delta m$ is proportional
to $\Lambda$ and it depends on the masses
of the heaviest fermions of the theory, the top and the bottom quark.
Here we will leave it as a free parameter.

The mass splitting gives rise to a mixing between the neutral component
of an isotriplet and the corresponding isosinglet.
This
mixing does not affect the calculation of $\Pi_{33}$ and $\Pi_{30}$
since in the corresponding graphs only charged particle loops appear.
On the contrary, for the calculation of $\Pi_{WW}$ one has to consider
the neutral mass eigenstates in the loops.

We have calculated the corrections to $\epsilon_1^{loop}$ coming from
the isospin violating PGB spectrum given in eq. (6.6-11), by using the
expression (5.5) in the context of
validity of the WSR's.

In Fig. 6 we show the SM radiative corrections for $\epsilon_1$ as a function
of $m_t$ (dotted line). They are compared
with the BESS predictions
at one-loop, which include the SM radiative corrections.
The band delimited by two solid (dashed) lines corresponds to
$\Delta m=200(300)~GeV$, and displays the complete range of variability in $z$.
In both cases the lower edge of the bands corresponds to $z=1$ while the
upper one is for $z\sim 0.3$. The figure is done for $M_V=1000$ GeV. As $M_V$
grows the bands shrink, keeping the lower edge
fixed because, in this case, only the PGB's contribute.

We see that the one-loop contribution is negative and, as discussed
for $\epsilon_3$, this is a general feature valid for any $SU(N)_L\otimes
SU(N)_R$ ($N>2$). In fact, as it appears from equation (5.9) the
total one-loop contribution arises from the sum of the quadruplets
contained in the adjoint representation of $SU(8)$, and each of this
contribution is negative definite. Therefore, also in this case,
by increasing $N$ the value of $\epsilon_1$ is decreased.

Fig. 7 is analogous to Fig. 6, but in this case, in the spirit of
ref. \cite{masse}, we take $\Delta m$ as a linear
function of $m_t$. The region delimited by solid lines corresponds to
$\Delta m=m_t$ and the one delimited by dashed lines to $\Delta m=1.5~m_t$.

In both figures the horizontal band corresponds to the $1 \sigma$ experimental
limit, as given in eq. (6.5). We note that the effect of the BESS correction is
to weaken the upper bound on the top mass.

\resection{Conclusions}

The main purpose of the work was the calculation of the corrections to the
self-energies of the gauge bosons of the standard model from the vector
and axial resonances and from the spin zero (pseudogoldstones) particles of
extended BESS. The corrections affect the scalar gauge boson self-energy terms
$\Pi_{WW}$, $\Pi_{33}$, $\Pi_{00}$, and $\Pi_{30}$ in the usual $SU(2)_L \times
U(1)$ notation. For convenience, to compare with the previous BESS results, we
have chosen the normalization in terms of the parameters $\eps_1$, $\eps_2$,
$\eps_3$, but any other parametrization would be equivalent.
\par
The mixings among vector and axial resonances and standard gauge bosons already
provide for a correction calculable at the tree-level, which turns out to be
the same as for $SU(2)$-BESS, that is essentially vanishing $\eps_1$ and
$\eps_2$, and an $\eps_3$ as in $SU(2)$-BESS. This holds for any $SU(N)_L
\times SU(N)_R$, the reason being that the new diagonal vector resonances
only mix with the electroweak $U(1)$ gauge boson.
\par
In addition to the self-energy corrections arising from the mixing, one has
loop contributions. At one loop, one has loops of Goldstones, loops of a
Goldstone and a vector resonance, or a Goldstone and an axial resonance, and
finally one has to add a Goldstone tadpole graph. The model provides for the
needed trilinear and quadrilinear couplings involving spin 1 and spin 0
particles. And, of course, one has also to take into account the possibilities
of mixings of the external $W^3$ and $Y$ legs with the vector and
axial-vector resonances.
\par
For the calculation of the one-loop contributions to $\eps_3$, which is isospin
symmetric, one can insert a common non-vanishing mass for all the
pseudogoldstones. The result one finds for $\eps_3$ goes back to that of
standard technicolor when the vector and axial-vector resonances are decoupled.
One again obtains this result when the vector and axial resonances are taken
as completely degenerate, due to their reciprocal cancellation in each case.
\par
The one-loop contributions to $\eps_1$ vanish for degenerate masses of the
$SU(2)$ pseudogoldstones multiplets, so that one has to take into account the
expected multiplet splittings. Fortunately the symmetry structure of the theory
implies sum rules for pseudogoldstones masses. A consequence of the sum rules
is also the cancellation of possible cut-off dependent terms. The total
$\eps_1$ is then obtained by summing on all quadruplets $P^{\pm}$, $P^o$, $P^3$
of pseudogoldstones taking their color multiplicity into account. The result at
one-loop is independent of the mass of the axial-vector resonances due to
cancellation between loops of pseudogoldstones and loops of axial-vectors.
\par
Finally, concerning $\eps_2$, one notices that in BESS there will be a
depression factor of the order $(M_W/M_V)^2$ with respect to $\eps_1$ so that
its role in the analysis is unimportant.
\par
Our quantitative analysis of $\eps_3$ and $\eps_1$ has to include certain
pseudogoldstone masses. For this we have used a calculation we had developed of
such masses, which includes the usual gauge contributions and in addition the
contributions from the interactions providing for ordinary fermion masses,
in the form of effective Yukawa couplings. The would-be massless goldstones
in absence of the yukawians, then acquire masses close to the heaviest
fermion, with consequent mass splittings reflecting the large splitting
between top and bottom quark.
\par
As far as the BESS parameters are concerned, we have reduced them by assuming
the validity of the Weinberg sum rules for the imaginary parts of the vector
and axial self-energies and saturating them with vector and axial-vector
dominance.
\par
The standard, QCD scaled technicolor, would correspond to an
additional specialization of the parameters contained in our set.
\par
The numerical analysis has of course to include electroweak radiative
corrections. Our assumption, physically plausible in the presence of new
degrees of freedom related to a larger scale, has been to take the usual one
loop radiative corrections of the standard model, by interpreting the Higgs
mass occurring there as the cut-off which regularizes BESS at high momenta.
\par
There exist experimental limits on the $\eps$'s from a set of observables
including $M_W/M_Z$, $A^{\ell}_{FB}$, and $\Gamma_{\ell}$, $A^{\tau}_{pol}$,
$A^{b}_{FB}$ and low energy weak data, which we have used to derive bounds on
the BESS parameters.
\par
We have illustrated in Fig.~4 the comparison with the present experimental
bounds on $\eps_3$ (the region inside the two horizontal lines is the 1$\sigma$
band). The theoretical $\eps_3$ versus the BESS parameter $z$ (which weights
the axial vector to vector coupling) is the solid line (the other parameters
are indicated in the caption, but their role is rather insensitive). One sees
that a definitive exclusion of this SU(8)-BESS is strictly yet not possible.
The one loop corrections have added a negative contribution bringing closer to
the experimental band. Indeed each SU(2) doublet of pseudogoldstones, vectors,
and axial vectors adds a negative contribution, so that by increasing $N$ in
the chiral structure $SU(N)_R\otimes SU(N)_L$ $\eps_3$ will further decrease.
\par
The comparison in terms of allowed region for the BESS parameters $M_V$ and $z$
(in Fig.~5) from the $\eps_3$ experimental limitations shows the region above
the solid line as the allowed one. The standard, QCD scaled, $SU(8)$
technicolor corresponds to the black dot. Inclusion of the one-loop corrections
has made this theory (which is however already afflicted by other serious
diseases of his own) marginally compatible, at least at present, with data.
\par
Also, in considering the strength of the conclusions for $SU(8)$-BESS, we
recall that in particular the validity of the second Weinberg sum rule has been
added within the assumptions, to limit the BESS parameters.
\par
The analysis for $\eps_1$ rests heavily on the isospin splittings within the
pseudogoldstones multiplets. We have employed a consistent parametrization for
them, supported by the theoretical analysis of their mass spectrum. The
horizontal  band in Fig.~6 is the experimental 1$\sigma$ band for $\eps_1$.
Depending on the top mass, the two theoretical bounds (solid lines and dashed
lines), corresponding to two different choices for the parameter characterizing
the pseudogoldstones mass splittings, have consistent overlaps with the
experimental band. By increasing $N$, in $SU(N)_R\otimes SU(N)_L$ more
quadruplets of pseudogoldstones arise, each giving a negative contribution to
$\eps_1$ at one loop, and the theoretical bands are increasingly lowered with
respect to the SM dotted-line in Fig.~6. Similar conclusions, qualitatively,
follows by explicitly exploiting the expected dependence of the splitting
parameter from the top mass, with the theoretical bounds taking a different
shape in this case. A common trend from these analyses of $\eps_1$ is that the
upper bounds, implied by $\eps_1$ on the top mass within the standard model,
get weakened in the presence of the strong electroweak sector.
\par
The careful calculations performed in this work in terms of extended BESS
continue with the program of transposing precise new experimental data into
bounds for a possible strong electroweak sector, in a rather general
formulation which tries to avoid reference to a particular new strong dynamics.
The fact that standard QCD-scaled $SU(8)$ technicolor now appears to be
marginally allowed by data, after a careful evaluation of the one loop
effects, should not be
interpreted as a possible revival for this model, which, as well known, has
always had deep theoretical difficulties. We consider that only by keeping
within a formulation as much as possible general, stressing symmetry aspects
and the main expected dynamical features, one can systematically follow the
increase in precision of the experimental data to circumscribe the theoretical
freedom still left for some type of new electroweak strong sector.

\newpage
\renewcommand{\theequation}{\Alph{section}.\arabic{equation}}
\appsectio
The generators of the $SU(8)$ algebra are
\be
T^A=(T^a,{\tilde T}^a,T_D,T_8^\alpha,T_8^{a\alpha},T_3^{\mu i},
{\bar T}_3^{\mu i})
\ee
with $A=1,\ldots ,63$, $a=1,2,3$, $\alpha =1,\ldots ,8$,
$\mu=(0,a)$ and $i=1,2,3$.
They are normalized by
\be
\Tr (T^A T^B)=\frac{1}{2}\delta^{AB}
\ee
We use the following representation
\be
T^a=\frac{1}{4}\left(
\begin{tabular}{c|c}\\
$\tau^a\otimes {\bf 1}_3$ & 0\\
\\
\hline \\
0& $\tau^a$\\\\
\end{tabular}\right)~~~~~
{\tilde T}^a=\frac{\sqrt{3}}{4}\left(\begin{tabular}{c|c}\\
$\tau^a\otimes {\bf 1}_3/3$ & 0\\
\\
\hline\\
0&$ -\tau^a$\\\\
\end{tabular}\right)
\ee
\be
T_D=\frac{\sqrt{3}}{4}\left(
\begin{tabular}{c|c}\\
${\bf 1}_2\otimes{\bf 1}_3/3$ & 0 \\
\\
\hline\\
0 & $-{\bf 1}_2$\\\\
\end{tabular}\right)
\ee
\be
T_8^\alpha=\frac{1}{\sqrt{8}}\left(
\begin{tabular}{c|c}\\
${\bf 1}_2\otimes \lambda^\alpha$ & 0 \\
\\
\hline\\
0 & 0\\\\
\end{tabular}\right)~~~~~
T_8^{a\alpha}=\frac{1}{\sqrt{8}}\left(\begin{tabular}{c|c}\\
$\tau^a\otimes \lambda^\alpha$ & 0\\
\\
\hline
\\
0&0\\\\
\end{tabular}\right)
\ee
\be
T_3^{\mu i}=\frac{1}{\sqrt{8}}\left(
\begin{tabular}{c|c}\\
0 & $\sigma_\mu\otimes\xi^i$\\
\\
\hline\\
$\sigma_\mu\otimes\xi^{iT}$ & 0\\\\
\end{tabular}\right)~~~~~
{\bar T}_3^{\mu i}=\frac{1}{\sqrt{8}}\left(
\begin{tabular}{c|c}\\
0 & $i \sigma_\mu\otimes\xi^i$\\
\\
\hline\\
$-i \sigma_\mu\otimes\xi^{iT}$ & 0\\\\
\end{tabular}\right)
\ee
where $\tau^a$ are the $SU(2)$ generators normalized by $\Tr(\tau^a\tau^b)=
2\delta^{ab}$, $\lambda^\alpha$ are the $SU(3)$ generators normalized by
$\Tr(\lambda^\alpha \lambda^\beta)=
2\delta^{\alpha\beta}$,
$\sigma^\mu\equiv({\bf 1}_2,\tau^a)$ and the $\xi^i$ are the three
orthogonal unit vectors in the three-dimensional vector space.

The commutation rules
\be
[T^A,T^B]= i f^{ABC}T^C
\ee
are obtained by using
\bea
& &\left [\tau^a,\tau^b \right ]=
2 i \epsilon^{abc}\tau^c\\
& &\lq\lambda^\alpha,\lambda^\beta\rq
=2 i  f^{\alpha\beta\gamma}\lambda^\gamma\\
& &\lq\lambda^\alpha,\lambda^\beta\rq_+
=\frac{4}{3}\delta^{\alpha\beta}+2 d^{\alpha\beta\gamma}\lambda^\gamma
\eea
We get the following non vanishing commutators:
\bea
\left[T^a,T^b\right]&=&\frac{i}{2}\epsilon^{abc} T^c \\
\left[T^a,{\tilde T}^b\right]&=&\frac{i}{2}\epsilon^{abc}{\tilde T}^c\\
\left[T^a,T_8^{b\alpha}\right] &=& \frac{i}{2}\epsilon^{abc}T_8^{c\alpha}\\
\lq T^a,T_3^{\mu i}\rq &=& \frac{i}{2}\delta^{\mu b}\epsilon^{abc}T_3^{ci}\\
\lq T^a,{\bar T}_3^{\mu i}\rq &=&
               \frac{i}{2}\delta^{\mu b}\epsilon^{abc}{\bar T}_3^{ci}\\
\left[{\tilde T}^a,{\tilde T}^b\right] &=&
                      \frac{i}{2}\epsilon^{abc} (T^c-\frac{2}{\sqrt{3}}
                                            {\tilde T}^c)\\
\left[{\tilde T}^a,T_8^{b\alpha}\right] &=& \frac{i}{2\sqrt{3}}
                        \epsilon^{abc} T_8^{c\alpha}\\
\lq{\tilde T}^a,T_3^{\mu i}\rq &=&
               -\delta^{\mu 0}\frac{i}{\sqrt{3}} {\bar T}_3^{a i}
               -\delta^{\mu a} \frac{i}{\sqrt{3}}{\bar T}_3^{0 i}
               -\delta^{\mu b} \frac{i}{2\sqrt{3}}
                        \epsilon^{abc} T_3^{c i}\\
\lq{\tilde T}^a,{\bar T}_3^{\mu i}\rq &=&
             \delta^{\mu 0}\frac{i}{\sqrt{3}}  T_3^{a i}
               +\delta^{\mu a} \frac{i}{\sqrt{3}} T_3^{0 i}
               -\delta^{\mu b} \frac{i}{2\sqrt{3}}
                        \epsilon^{abc} {\bar T}_3^{c i}\\
\lq T_D,T_3^{\mu i}\rq &=& -\frac{i}{\sqrt{3}}
                       {\bar T}_3^{\mu i}\\
\lq T_D,{\bar T}_3^{\mu i}\rq &=& \frac{i}{\sqrt{3}}
                        T_3^{\mu i}\\
\lq T_8^\alpha,T_8^\beta\rq &=& \frac{i}{\sqrt{2}} f^{\alpha\beta\gamma}
                       T_8^\gamma\\
\lq T_8^\alpha,T_8^{a\beta}\rq &=& \frac{i}{\sqrt{2}}
            f^{\alpha\beta\gamma} T_8^{a\gamma}\\
\lq T_8^\alpha,T_3^{\mu i}\rq &=& -\frac{1}{4\sqrt{2}}
       \lq[(\lambda^\alpha)_{im}-(\lambda^\alpha)_{mi}] T_3^{\mu m}+ i
       [(\lambda^\alpha)_{im}+(\lambda^\alpha)_{mi}] {\bar T}_3^{\mu m}\rq\\
\lq T_8^\alpha,{\bar T}_3^{\mu i}\rq &=& -\frac{1}{4\sqrt{2}}
       \lq[(\lambda^\alpha)_{im}-(\lambda^\alpha)_{mi}] {\bar T}_3^{\mu m}- i
       [(\lambda^\alpha)_{im}+(\lambda^\alpha)_{mi}] T_3^{\mu m}\rq\\
\lq T_8^{a\alpha},T_8^{b\beta}\rq &=&\frac{i}{2}\delta^{\alpha\beta}
            \epsilon^{abc}\lq T^c+\frac{1}{\sqrt{3}} {\tilde T}^c\rq+
                    \frac{i}{\sqrt{2}}\lq
           \delta^{ab}  f^{\alpha\beta\gamma}T_8^\gamma
             +\epsilon^{abc}d^{\alpha\beta\gamma}T_8^{c\gamma}\rq\\
\lq T_8^{a\alpha},T_3^{\mu i}\rq &=& -\frac{1}{4\sqrt{2}}
       [(\lambda^\alpha)_{im}-(\lambda^\alpha)_{mi}]
           [\delta^{\mu 0}T_3^{a m}+ \delta^{\mu b}
         \epsilon^{abc}{\bar T}_3^{cm}+
             \delta^{\mu a}T_3^{0 m}]\nn\\
      & &-\frac{i}{4\sqrt{2}}
       [(\lambda^\alpha)_{im}+(\lambda^\alpha)_{mi}]
           [\delta^{\mu 0}{\bar T}_3^{a m}-\delta^{\mu b}
             \epsilon^{abc}T_3^{cm}+
             \delta^{\mu a}{\bar T}_3^{0 m}]\\
\lq T_8^{a\alpha},{\bar T}_3^{\mu i}\rq &=& -\frac{1}{4\sqrt{2}}
       [(\lambda^\alpha)_{im}-(\lambda^\alpha)_{mi}]
           [\delta^{\mu 0}{\bar T}_3^{a m}-\delta^{\mu b}
          \epsilon^{abc}T_3^{cm}+
             \delta^{\mu a}{\bar T}_3^{0 m}]\nn\\
      & &+\frac{i}{4\sqrt{2}}
       [(\lambda^\alpha)_{im}+(\lambda^\alpha)_{mi}]
           [\delta^{\mu 0} T_3^{a m}+\delta^{\mu b}
              \epsilon^{abc}{\bar T}_3^{cm}+
             \delta^{\mu a}T_3^{0 m}]\\
\lq T_3^{\mu i},T_3^{\nu j}\rq &=&
\lq {\bar T}_3^{\mu i},{\bar T}_3^{\nu j}\rq~=~\frac{i}{2}
\delta^{ij}\delta^{\mu a}\delta^{\nu b}\epsilon^{abc}
\lq T^c-\frac{1}{\sqrt{3}}{\tilde T}^c\rq+~color~terms\\
\lq T_3^{\mu i},{\bar T}_3^{\nu j}\rq &=&
-\frac{i}{\sqrt{3}}\delta^{ij}\lq
\delta^{\mu \nu} T_D+(\delta^{\mu 0}\delta^{\nu a}+\delta^{\nu 0}
\delta^{\mu a})
{\tilde T}^a\rq+~color~terms
\eea

\appsection

Let us consider the mixing in the $SU(2)_L\otimes U(1)_R$ sector.

{}From ${\cal L}^{(2)}$ given in eq. (2.27) one can easily obtain the
squared mass matrix, ${\bf M}^2$,
 of the neutral sector in the basis $(Y,W^3,V^3,V_D,A^3)$.
By performing the transformation
${\cal M}={\cal U}_2 {\cal U}_1 {\bf M}^2 {\cal U}_1^T
{\cal U}_2^T$ with  ${\cal U}_{1,2}$ orthogonal matrices given by
\be
{\cal U}_1=
\left (\begin{array}{ccccc}
c_\theta & s_\theta & 0 & 0 & 0\\
-s_\theta & c_\theta & 0 & 0 & 0\\
0 & 0 & 1 & 0 & 0\\
0 & 0 & 0 & 1 & 0\\
0 & 0 & 0 & 0 & 1
\end{array}\right)~~
{\cal U}_2=
\left (\begin{array}{ccccc}
c_\psi c_\eta & 0 & s_\psi & c_\psi s_\eta & 0\\
0 & 1 & 0 & 0 & 0\\
-s_\psi c_\eta & 0 & c_\psi & -s_\psi s_\eta & 0\\
-s_\eta & 0 & 0 & c_\eta & 0\\
0 & 0 & 0 & 0 & 1
\end{array}\right)
\ee
and $\psi\simeq 2 x s_\theta $, $\eta\simeq 2 x y  s_\theta $,
we get
\be
{\cal M}=
\left (\begin{array}{cc}
0 & 0\\
0 & {\bar{\cal M}}
\end{array}\right )
\ee
with
\be
{\bar {\cal M}}=\frac{v^2}{4} G^2
\left (\begin{array}{cccc}
 1+\dd{\frac{E_V^2}{R_V}}+\dd{\frac{E_0^2}{R_0}}+\dd{\frac{E_A^2}{R_A}}&
 -\dd{\frac{E_V}{R_V}} &\dd{\frac{E_0}{R_0}}-\dd{\frac{E_1}{R_V}} &
\dd{\frac{E_A}{R_A}}\\
-\dd{\frac{E_V}{R_V}}& \dd{\frac{1}{R_V}} & \dd{\frac{E_1}{E_V R_V}} & 0\\
\dd{\frac{E_0}{R_0}}-\dd{\frac{E_1}{R_V}} &
\dd{\frac{E_1}{E_V R_V}} & \dd{\frac{1}{R_0}}+\dd{\frac{E_1^2}{E_V^2 R_V}}
& 0\\
\dd{\frac{E_A}{R_A}} & 0 & 0 & \dd{\frac{1}{R_A}}
\end{array}\right)
\ee
where
$$
E_V = \frac{c_{2\theta}}{c_\theta} N x~~~~
E_A = \frac{z}{c_\theta} x~~~~
E_0 = 2 y\frac{s_\theta^2}{c_\theta}  B N x ~~~~
E_1 = 4 y\frac{c_{2\theta} s^2_\theta}{c_\theta}  \frac{N}{B} x^3\nn
$$
\be
R_V = \frac{r_V}{c_\theta^2} N^2~~~ R_A=\frac{r_A}{c_\theta^2}~~~
R_0 = \frac{r_V}{c_\theta^2} B^2 N^2
\ee
with $r_V$, $r_A$, $z$ defined in eq. (2.33), and
\bea
B^2 &=& 1+4 x^2 s^2_\theta (1+y^2)\\
B^2 N^2 &=& 1+4 x^2 s^2_\theta y^2
\eea
One recovers the $SU(2)$-BESS model by putting $y=0$ (corresponding to the
decoupling of $V_D$).
The eigenvalues of ${\bar {\cal M}}$
are the squared masses of the $Z$, $V^3$, $V_D$ and $A^3$ bosons, and,
in the
large $\gs$ limit (that is
for small $x$) are the following:
\bea
\lambda_1 &\simeq& \frac{v^2}{4} G^2 \left(1-\frac{E^2_V}{1-R_V}-
                \frac{E^2_0}{1-R_0}-\frac{E^2_A}{1-R_A}
                \right)\simeq M^2_Z\\
\lambda_2 &\simeq& \frac{v^2}{4} \frac{G^2}{ R_V}\simeq M^2_{V^3}\\
\lambda_3 &\simeq& \frac{v^2}{4} \frac{G^2}{R_0}\simeq M^2_{V_D}\\
\lambda_4 &\simeq& \frac{v^2}{4} \frac{G^2}{ R_A}\simeq M^2_{A^3}
\eea
and substituting the expressions of eqs. (B.4)-(B.6) we obtain
the values
given in eqs. (2.30)-(2.32).

\appsection

We use the notation
\be
\pi^A T^A=\pi^a T^a+{\tilde \pi}^a {\tilde T}^a+\pi_D T_D+
\pi_8^{\alpha} T_8^{\alpha}
        +\pi_8^{\alpha a} T_8^{\alpha a}+\pi_3^{\mu i} T_3^{\mu i}+
{\bar \pi}_3^{\mu i} {\bar T}_3^{\mu i}
\ee
and analogously for $V^A T^A$.
It turns out that $\pi_3^{\mu i}$ and ${\bar \pi}_3^{\mu i}$ do not have the
right transformation properties under the color $SU(3)$. Therefore we
introduce the linear combinations:
\be
P_3^{\mu i}=\frac{\pi_3^{\mu i}- i {\bar \pi}_3^{\mu i}}{\sqrt{2}}~~~~~
{\bar P}_3^{\mu i}=\frac{\pi_3^{\mu i}+ i {\bar \pi}_3^{\mu i}}{\sqrt{2}}
\ee
The charged components of the $SU(2)$ triplets are
defined in the standard way:
\be
P_3^{\pm i}=\frac{P_3^{1 i}\pm i P_3^{2 i}}{\sqrt{2}}
\ee

Notice that
$(P_3^{\mu i})^\dagger={\bar P}_3^{\mu i}$.

In an analogous way for the $V$ and $A$ gauge bosons we introduce the
following combinations:
\bea
{\cal V}_3^{\mu i}&=&\frac{V_3^{\mu i}- i {\bar V}_3^{\mu i}}{\sqrt{2}}~~~~~
{\bar {\cal V}}_3^{\mu i}=\frac{V_3^{\mu i}+ i {\bar V}_3^{\mu i}}{\sqrt{2}}\\
{\cal A}_3^{\mu i}&=&\frac{A_3^{\mu i}- i {\bar A}_3^{\mu i}}{\sqrt{2}}~~~~~
{\bar {\cal A}}_3^{\mu i}=\frac{A_3^{\mu i}+ i {\bar A}_3^{\mu i}}{\sqrt{2}}
\eea
and again
\be
{\cal V}_3^{\pm i}=\frac{{\cal V}_3^{1 i}\pm i{\cal V}_3^{2 i}}
{\sqrt{2}}~~~~~~~~
{\cal A}_3^{\pm i}=\frac{{\cal A}_3^{1 i}\pm i{\cal A}_3^{2 i}}{\sqrt{2}}
\ee

In the following table we list the 63 Goldstone bosons with their quantum
numbers and transformation properties under $SU(2)_L$ and $SU(3)_c$ (here
$Y=2(Q-T^3)$ is the hypercharge):

\begin{center}
\begin{tabular}{|c|c|c|c|c|}
\hline
&$SU(2)_L$&$SU(3)_c$&$Q$&$Y$\\
\hline
$\pi^+~({\tilde\pi^+})$ & & & 1 &\\
$\pi^-~({\tilde\pi^-})$ &3 &1 &-1 &0\\
$\pi^3~({\tilde\pi^3})$ & & & 0 &\\
\hline
$\pi_D $&1 &1 &0 &0\\
\hline
$\pi_8^\alpha$ &1 &8 &0 &0\\
\hline
$\pi_8^{\alpha +} $& & & 1 &\\
$\pi_8^{\alpha -}$ &3 &8 &-1 &0\\
$\pi_8^{\alpha 3}$ & & & 0 &\\
\hline
  & & & &\\
$P_3^{0i}~({\bar P}_3^{0i})$ & 1&3&$\frac{2}{3}~(-\frac{2}{3})$& \\
 & & & &\\
\cline{1-4}
& & & & \\
$P_3^{+i}~({\bar P}_3^{+i})$& & &$\frac{5}{3}~(-\frac{5}{3})$
                              &$\frac{4}{3}~(-\frac{4}{3})$\\
$P_3^{-i}~({\bar P}_3^{-i})$ & 3&3&$-\frac{1}{3}~(\frac{1}{3})$& \\
$P_3^{3i}~({\bar P}_3^{3i})$ & & &$\frac{2}{3}~(-\frac{2}{3})$& \\
& & & &\\
\hline
\end{tabular}
\end{center}

\appsection

Because of the mixing among the ordinary gauge vector bosons
$W^3$, $Y$ and $V^3$, $A^3$,
 the parameters $\epsilon_i$ will in general receive
a contribution from the self-energies of the new vector bosons $V^3$ and
$A^3$.
To evaluate such contributions one can first compute the imaginary parts of
the
$V^3$ and $A^3$ self-energies and then insert them in appropriate dispersion
relations. We have obtained:
\be
\delta\epsilon_3^V=\left(\frac {g}{g''}\right)^2
\left\{
\frac {1}{\pi}
\int_{4m_\Pi^2}^{\infty} \frac {ds}{s^2}
\frac
{M_V^4 F_V(s)}
{(s-M_V^2)^2+ F_V^2(s)}
-1\right\}
\ee
In the previous formula $m_\Pi$ stands for the common mass of the PGB's
and
\bea
F_V(s)&=& s\frac{\Gamma_V}{M_V}
 \frac{ \left(1-\displaystyle\frac {4m_\Pi^2}{s}\right)^{3/2}}
{ \left(1-\displaystyle\frac {4m_\Pi^2}{M_V^2}\right)^{3/2}}\\
\Gamma_V&=&\frac{G_F^2}{24\pi}M_V^5\frac{(1-z^2)^2}{\gs^2}N_d^2\left(
1-\frac{4m_\Pi^2}{M_V^2}\right)^{3/2}
\eea
In eq. (D.1), $\delta\epsilon_3^V$ is the approximate contribution to
$\epsilon_3$ due to the self-energy of the $V^3$ vector boson. It has been
derived by including just the loops from the $(N_d^2-1)$ diagrams with
internal PGB's,
and neglecting loops with internal $V$ and $A$ fields.
We have approximately taken into account the diagram with a loop of $W's$
by replacing $N_d^2-1$ with $N_d^2$ in eq. (D.3).

In a similar way we have obtained the contribution to $\epsilon_3$ due to the
self-energy of the axial-vector boson $A^3$:
\be
\delta\epsilon_3^A=-\left(\frac {g}{g''}\right)^2 z^2
\left\{
\frac {1}{\pi}
\int_{(m_\Pi+M_V)^2}^{\infty} \frac {ds}{s^2}
\frac
{M_A^4 F_A(s)}
{(s-M_A^2)^2+ F_A^2(s)}-
1\right\}
\ee
with
\bea
F_A(s)&=&M_A \Gamma_A  \displaystyle
\frac{\left(3+\displaystyle\frac {s}{4M_V^2}f_s\right)}
{\left(3+\displaystyle\frac {M_A^2}{4M_V^2}f\right)}
\sqrt{\displaystyle\frac {f_s} {f}}\\
\Gamma_A&=&\frac{\sqrt{2} G_F M_A^3}{48\pi}\left(1-\frac{M_V^2}{M_A^2}\right)^2
N_d^2 z^2\left(3+\frac{M_A^2}{4M_V^2}f\right)\sqrt{f}\\
f&=&1+\frac{m_\Pi^4}{M_A^4}+\frac{M_V^4}{M_A^4}-2\frac{m_\Pi^2
M_V^2}{M_A^4}-2\frac
{m_\Pi^2}{M_A^2}-2\frac{M_V^2}{M_A^2}\\
f_s&=&1+\frac{m_\Pi^4}{s^2}+\frac{M_V^4}{s^2}-2\frac{m_\Pi^2 M_V^2}{s^2}-2\frac
{m_\Pi^2}{s}-2\frac{M_V^2}{s}
\eea
In eq. (D.4) the $A^3$ self-energy includes just the contribution from
one-loop diagrams with an internal $(\pi,V)$ pair.

By imposing the WSR's (eq. (6.2)), from the previous equations one obtains
an expression for $\delta\epsilon_3=\delta\epsilon_3^V+\delta\epsilon_3^A$
depending on $M_V$ and $z$.

\newpage
\begin{center}
  \begin{Large}
  \begin{bf}
  Figures
  \end{bf}
  \end{Large}
  \end{center}
  \vspace{5mm}
%%%%%%%%%%%%
\begin{figure}[h]
\begin{center}
\includegraphics[width=9cm,angle=0]{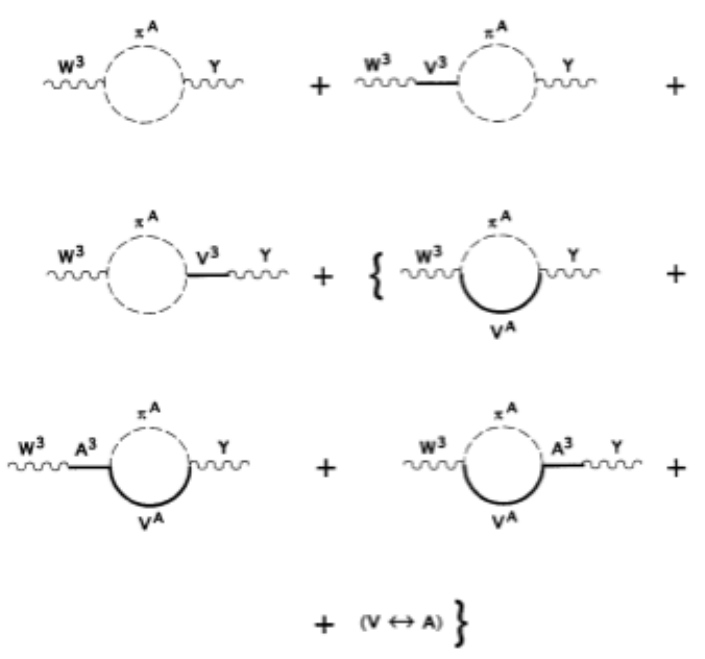}
\caption{{\it Graphs included in the calculation of $\epsilon_3^{loop}$ of eq. (4.1).}} 
\end{center}
\end{figure}
%%%%%%%%%%%%

%%%%%%%%%%%%
\begin{figure}[h]
\begin{center}
\includegraphics[width=9cm,angle=0]{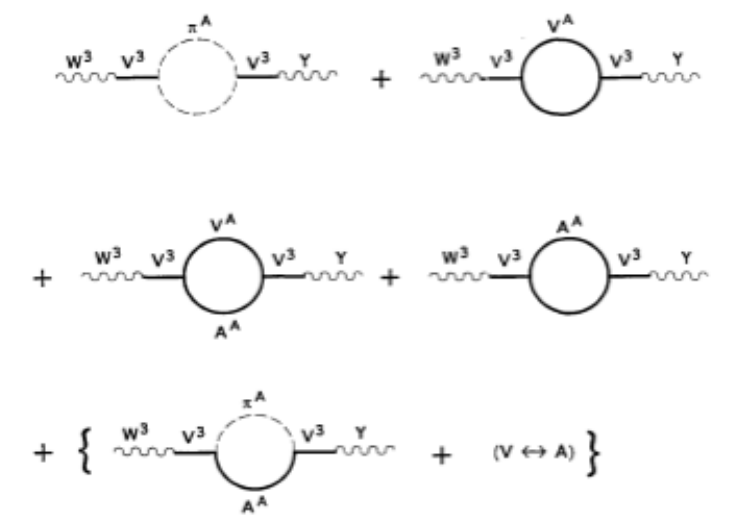}
\caption{{\it Corrections to $\epsilon_3^{loop}$ coming from the $V^3$
               and $A^3$ self-energy contributions.}} 
\end{center}
\end{figure}
%%%%%%%%%%%%

%%%%%%%%%%%%
\begin{figure}[h]
\begin{center}
\includegraphics[width=9cm,angle=0]{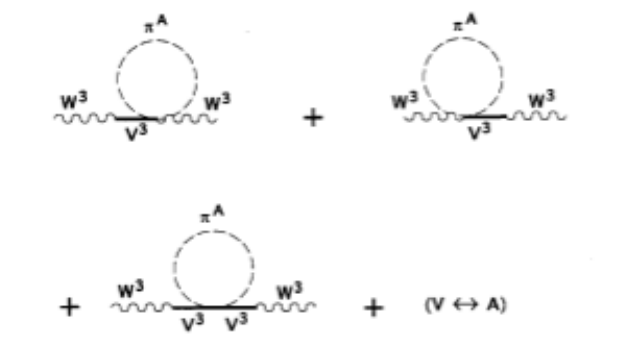}
\caption{{\it Tadpole graphs contributing to $\Pi_{33}$ which are relevant
               for the calculation of $\epsilon_1^{loop}$.}} 
\end{center}
\end{figure}
%%%%%%%%%%%%

%%%%%%%%%%%%
\begin{figure}[h]
\begin{center}
\includegraphics[width=9cm,angle=0]{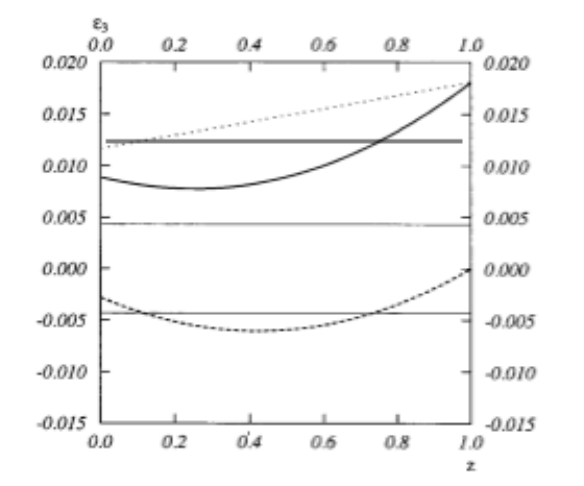}
\caption{{\it $\epsilon_3$ versus $z$ for $\Lambda=1.5~TeV$, $M_V=1~TeV$
           	and $m_\Pi/\Lambda=0.35$. The dotted line is the tree-level
        	BESS model contribution plus the SM radiative corrections,
		 the dashed
		line is the one-loop contribution, and
		the solid line is the total correction. Also shown is the
		experimental 1$\sigma$ band.}} 
\end{center}
\end{figure}
%%%%%%%%%%%%

%%%%%%%%%%%%
\begin{figure}[h]
\begin{center}
\includegraphics[width=9cm,angle=0]{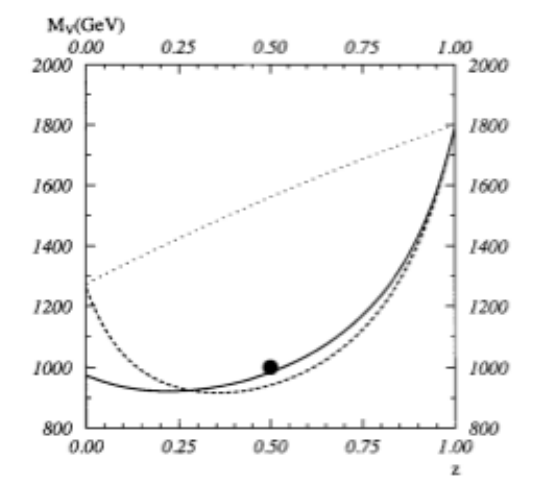}
\caption{{\it 90\% C.L. allowed region in the plane $(z,M_V)$ from
        	$\epsilon_3$ for $\Lambda=1.5~TeV$.
	        The dotted line is the lower bound on $M_V$ coming from the
		tree-level BESS model contribution plus the SM radiative
		corrections, the solid (dashed) line is the bound coming from
		the total one-loop effect for $m_\Pi/\Lambda=0.35$
                ($m_\Pi/\Lambda=0.10$).
           	The black dot shows the case of technicolor with one family
		of technifermions and $N_{TC}=3$.}} 
\end{center}
\end{figure}
%%%%%%%%%%%%

%%%%%%%%%%%%
\begin{figure}[h]
\begin{center}
\includegraphics[width=9cm,angle=0]{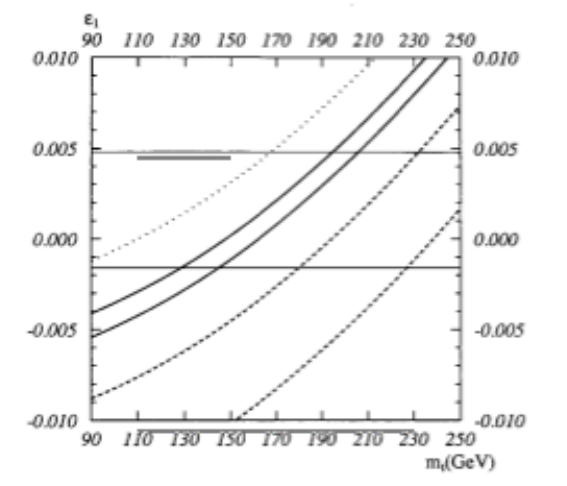}
\caption{{\it $\epsilon_1$ versus $m_t$ for $\Lambda=1~TeV$ and $M_V=1~TeV$.
           	The dotted line corresponds to the SM radiative corrections.
		The band delimited by two solid (dashed)
		lines corresponds to $\Delta m=200~GeV$
                ($\Delta m=300~GeV$) and displays
		the complete range of variability in $z$ ($0\le z\le 1$).
		Also shown is the experimental 1$\sigma$ band.}} 
\end{center}
\end{figure}
%%%%%%%%%%%%

%%%%%%%%%%%%
\begin{figure}[h]
\begin{center}
\includegraphics[width=9cm,angle=0]{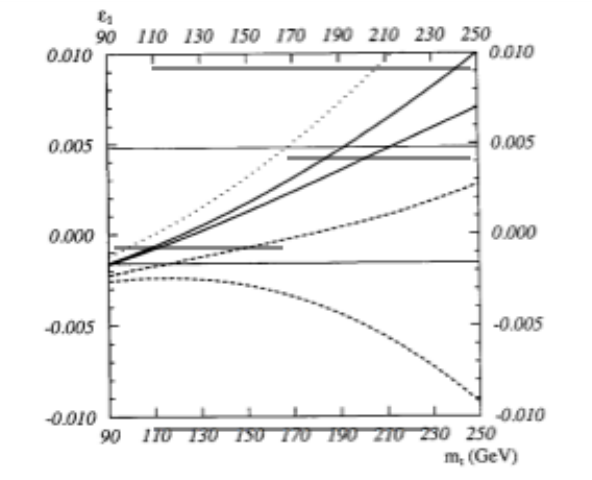}
\caption{{\it $\epsilon_1$ versus $m_t$ for $\Lambda=1~TeV$ and $M_V=1~TeV$.
           	The dotted line corresponds to the SM radiative corrections.
		The band delimited by two solid (dashed)
		lines corresponds to $\Delta m=m_t$
                ($\Delta m = 1.5~m_t$) and displays
		the complete range of variability in $z$ ($0\le z\le 1$).
		Also shown is the experimental 1$\sigma$ band.}} 
\end{center}
\end{figure}
%%%%%%%%%%%%

\end{document}